\documentclass[runningheads]{llncs}

\usepackage{iciap}

\usepackage{iciapabbrv}

\usepackage{graphicx}
\usepackage{booktabs}
\usepackage{float}

\usepackage[accsupp]{axessibility}  

\usepackage{hyperref}

\usepackage{orcidlink}

\begin{document}

\title{DeepFRI Demystified: Interpretability vs. Accuracy in AI Protein Function Prediction} 

\author{Ananya Krishna\inst{1} \and
Valentina Simon\inst{2} \and
Arjan Kohli\inst{1}}

\authorrunning{A. Krishna et al.}

\institute{Yale University, New Haven CT, 06510, USA
\and
Microsoft, Redmond WA, 98052, USA.}

\maketitle

\begin{abstract}
Machine learning technologies for protein function prediction are black box models. Despite their potential to identify key drug targets with high accuracy and accelerate therapy development, the adoption of these methods depends on verifying their findings. This study evaluates DeepFRI, a leading Graph Convolutional Network (GCN)-based tool, using advanced explainability techniques—GradCAM, Excitation Backpropagation, and PGExplainer—and adversarial robustness tests. Our findings reveal that the model's predictions often prioritize conserved motifs over truly deterministic residues, complicating the identification of functional sites. Quantitative analyses show that explainability methods differ significantly in granularity, with GradCAM providing broad relevance and PGExplainer pinpointing specific active sites. These results highlight trade-offs between accuracy and interpretability, suggesting areas for improvement in DeepFRI's architecture to enhance its trustworthiness in drug discovery and regulatory settings.
\end{abstract}

\section{Introduction}
\label{sec:intro}

Explainability in artificial intelligence (XAI) is pivotal for adopting and effectively using AI in healthcare and biotechnology \cite{biswas2024explainable}. A Deloitte survey found that 75\% of professionals in these fields consider explainability essential for AI adoption \cite{deloitte2019ai}. Additionally, about 90\% of biopharmaceutical companies require explainable AI to comply with regulatory standards from agencies like the FDA and EMA, which increasingly mandate transparent AI models for drug development and approval \cite{ibm2024responsibleai}.
Despite breakthroughs in protein structure prediction—most notably AlphaFold—functional annotation tools still lag in both performance and trustworthiness. Of existing methods, DeepFRI \cite{gligorijevic2021structure} is the most comprehensive: it uses graph convolutional networks trained on PDB‐derived structural graphs to assign Gene Ontology (GO) terms. Alternative approaches such as ContactPFP \cite{kagaya2022contactpfp}, which relies on residue–residue contact maps, and 3D CNNs \cite{torng2019amino}, which identify functional pockets, often miss crucial topological context. Homology‐independent tools like FFPred 3 \cite{cozzetto2016ffpred} (SVM‐based) and multimodal frameworks like DeepGO \cite{kulmanov2018deepgo} offer complementary strengths but still lack end-to-end interpretability. Although DeepFRI achieves state-of-the-art performance, its explainability and robustness beyond preliminary GradCAM analyses remain unexplored. Here, we apply a suite of post hoc explainability techniques to DeepFRI to pinpoint the most influential amino acid residues driving its predictions, uncover potential vulnerabilities, and enhance the transparency of protein function models \cite{pope2019explainability}.

\section{Methods}
We evaluate the DeepFRI model \cite{gligorijevic2021structure}, which predicts a protein's function from its structure. Our analysis identifies which structural motifs are most influential to the model's predictions across protein test cases. The creators of DeepFRI utilized GradCAM \cite{pope2019explainability} for a qualititative understanding of explainability. We build upon this by reproducing their GradCAM results, corroborating these findings with other post-hoc explainability techniques (Excitation Backpropagation and PG Explainer \cite{pope2019explainability}), and introducing quantitative scores like sparsity \cite{pope2019explainability}. 

We randomly selected a diverse set of 124 binding proteins and fed their native FASTA sequences to DeepFRI. We verified that none of the proteins were in DeepFRI's training set. We focus on binding proteins due to their clear active sites and distinguishable functional residues. Prioritizing slightly different features of the model's architecture in calculation, GradCAM, Excitation Backpropagation, and PG Explainer all produce saliency heat-maps that highlight the most relevant/deterministic input regions (ie. amino acids in the sequence) prioritized by DeepFRI for prediction. These amino acids correlate with the highest activations in the GCN.

We also apply a modified version of DeepFool \cite{moosavi2016deepfool} to understand how changes in sequence information change DeepFRI's behavior. Similar to DeepFool's emphasis on iteratively perturbing a minimal number of pixels in an image such that the majority of the original input is preserved, we identify the smallest number of mutations necessary to cause DeepFRI to misclassify a protein's function. After calculating the activation scores for each amino acid residue with GradCAM, we assign a directly proportional probability score to each residue in the sequence. Using hardcoded "dissimilar amino acids", a FASTA sequence is iteratively mutated. Only one residue is mutated at a time, with previous mutations preserved from prior iterations, before the sequence fed to the model for prediction. This iterative process is conducted 50 times to produce a plot that showcases the average "mutation threshold" for a particular prediction, or the number of point mutations necessary to cause a misclassification. This enables us to understand the model's sensitivity and generalization. 

\section{Results and Discussion}
We have results for thousands of combinations of explainability methods, proteins, and GO terms, but we outline two representative examples here in detail.

\subsection{Case Study 1: Lac Repressor Protein}
The lac repressor is a DNA-binding protein that inhibits transcription of lactose metabolism genes when sugar is absent \cite{lewis2006lacrepressor}. We used GradCAM to identify regions relevant to DeepFRI's predictions.

\begin{figure} [H]
    \centering
    \includegraphics[width=0.7\linewidth]{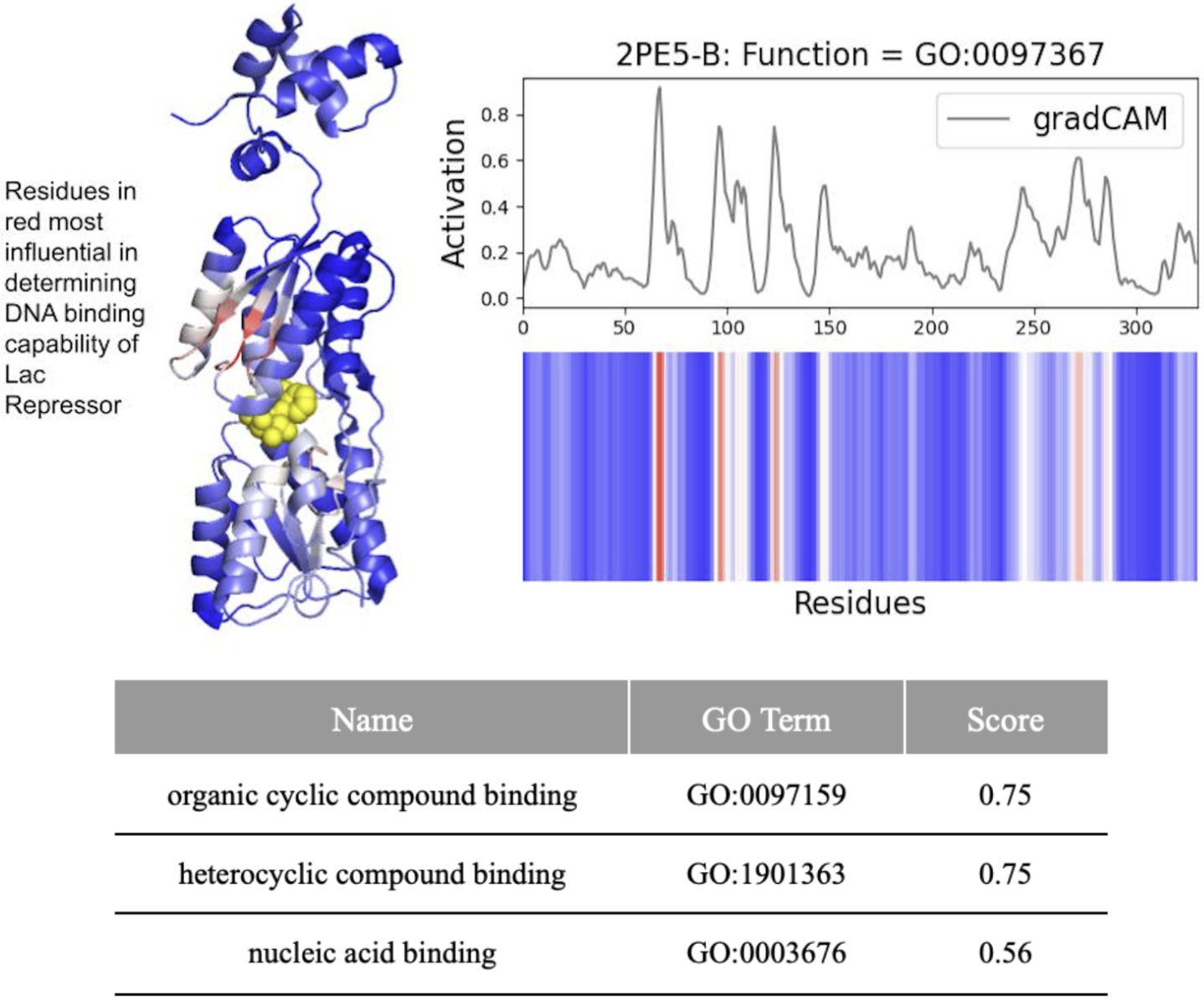}
    \caption{Above: GradCAM on 2PE5B, the Lac Repressor. The line graph shows neural network activations plotted for residues from beginning to end. The residues highlighted in red (60-65, 97-99, 124-6, 275) are the most important in DeepFRI's prediction of the function of the Lac Repressor. Yellow spheres represent water and ligands. Below: Table with DeepFRI's top three functional prediction for the Lac Repressor with confidence scores. All of these functions are consistent with the true function of the lac repressor.}
    \label{fig:enter-label}
\end{figure}

Shown in figure 1, the red/most-relevant residues correspond to a distinctive feature of the lac repressor, the singular beta pleated sheet found in its monomeric subunits. This sheet stabilizes the folded structure, but is surprisingly not related to the protein's actual function. We found DeepFRI prioritized amino acids that control protein structure over protein function in over half the proteins we examined. This could result from DeepFRI being trained on homology data, a critical metric for determining biological function, but not always for determining functional residues. This exemplifies the tradeoff between accuracy and interpretability.

Next, we investigated the robustness of DeepFRI with DeepFool, initially testing a baseline of two manually altered sequences. First, a completely scrambled version of the sequence in which all amino acids had been rearranged. This, as expected, yielded vastly different GO classifications related to metabolism. Second, we inputted a sequence in which only the red-labeled residues (the most statistically significantly relevant ones) were replaced with their least similar amino acid counterparts. Surprisingly, DeepFRI correctly predicted the same GO classifications as the native sequence, with the same prior confidence.

When amino acids were mutated from most to least relevant (according to GradCAM), it took 206 mutations of the 330 amino acid lac repressor for misclassification to occur. However, when 50 iterations of stochastic mutations (sampling from a probability distribution of residues weighted by relevance) was performed, the mutation threshold approached 75.

The fact that the model is more sensitive to mutations that are not the "most relevant" poses questions about the true relevance of each residue. It is highly possible that the amino acids prioritized in predictions do not have significant differences in relative weight, as supported by the earlier sparsity findings for GradCAM. This extreme robustness—needing to mutate over two‐thirds of residues before the GCN’s output flips—suggests that the model’s message‐passing is dominated by stable local neighborhoods. In other words, unless enough residues are changed to reconfigure those “neighborhood” features en masse, the graph convolution layers keep aggregating very similar structural signals, so the prediction remains the same. While is DeepFRI probably more reliable against spurious noise,it may miss subtle, biologically meaningful alterations (e.g. an active-site mutation) unless a large fraction of the structure is disrupted. The model is also probably biased towards local structural context: a large number of subgraphs must break to force the network to seek a new neighborhood pattern and its receptive field is too coarse to pick up on smaller, long-range interactions.

\begin{figure}[H]
    \centering
    \includegraphics[width=0.42\linewidth]{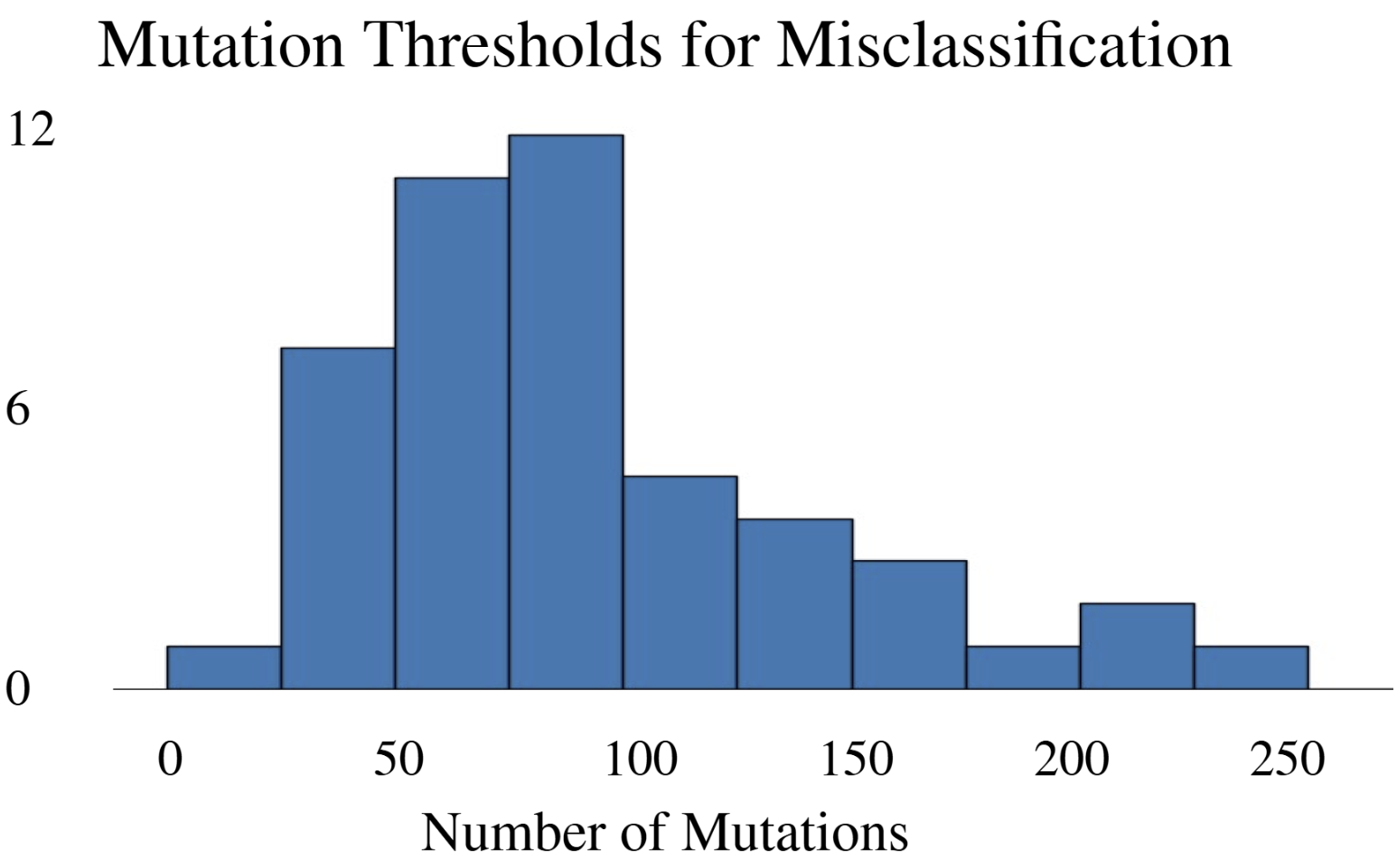}
    \caption[width=2\textwidth]{Mutations necessary to incorrectly predict protein function (50 trials) }
    \label{fig:example}
\end{figure}

\subsection{Case Study 2: ARF6 GTPase}

2W83-E, or ARF6 GTPase, regulates intracellular transport and movement of proteins. It promotes pathological processes related to vascular instability, tumor formation, growth, and metastasis \cite{grossmann2019arf6}. With this protein, we compare our three explainability methods and our quantitative metric sparsity.

 \begin{figure}[H]
  \centering
  \includegraphics[width=0.8\linewidth]{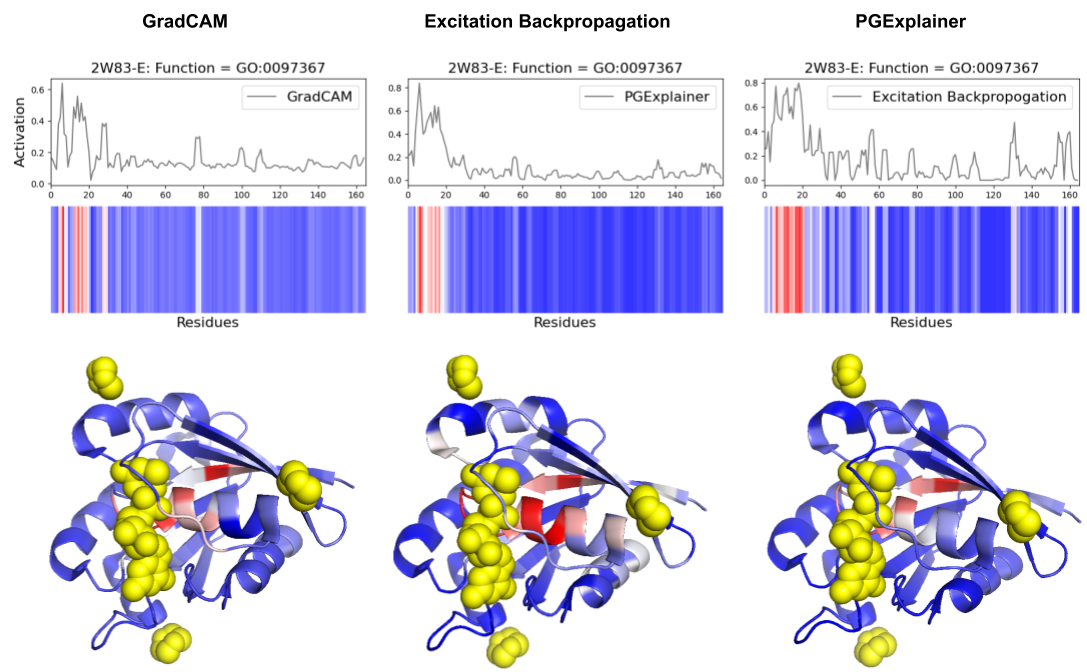}
  \caption{Comparison of Three Methods: 
  GradCAM, Excitation Backpropagation, and PGExplainer calculated for GTPase  (2W83-E). Figure highlights relevant regions (red, higher activation) over irrelevant regions (blue, lower activation) in prediction/likelihood of carbohydrate binding. (GO id: 0097367)}
  \label{fig:example}
\end{figure}

As shown in figure 3, the fact that GradCAM, Excitation Backpropagation (EB), and PGExplainer all zero in on the P-loop residues (0–20) in ARF6 GTPase gives us strong cross-validation of our post-hoc methods. GradCAM’s broad, graded heatmap shows that the model examines wide swath of structural context; EB’s binary masks give a clear include/exclude signal; and PGExplainer’s perturbation-based scores pinpoint individual catalytic residues. The P loop implicated by all three is a highly conserved structure that binds to the phosphate groups of GTP. Taken together, the methods build confidence that DeepFRI is focusing on genuinely biochemically meaningful motifs, and they illustrate how gradient- and perturbation-based techniques complement one another to reveal different facets of model reasoning.

We also determined sparsity (see fig. 4) for each method, or how many amino acids were used to predict protein functionality. GradCAM is the least sparse method, with a 3-fold smaller mean sparsity than PG Explainer and 17-fold smaller than Excitation Backpropagation. GradCAM’s low sparsity implies robustness (it won’t flip its view for small changes) but sacrifices localization precision. EB’s highly sparse, binary maps are immediately interpretable—ideal when seeking a clear yes/no on residue importance but may miss subtler contributions. PGExplainer sits between these extremes, offering per-residue relevance scores that can guide both narrow hotspot identification and broader structural insights. As a toolkit, users can choose the method that best fits their goals, whether that’s robust screening, precise mutagenesis guidance, or detailed structural analysis.
 
\begin{figure}[H]
  \centering

  \begin{minipage}[t]{0.32\linewidth}
    \centering
    \includegraphics[width=\linewidth]{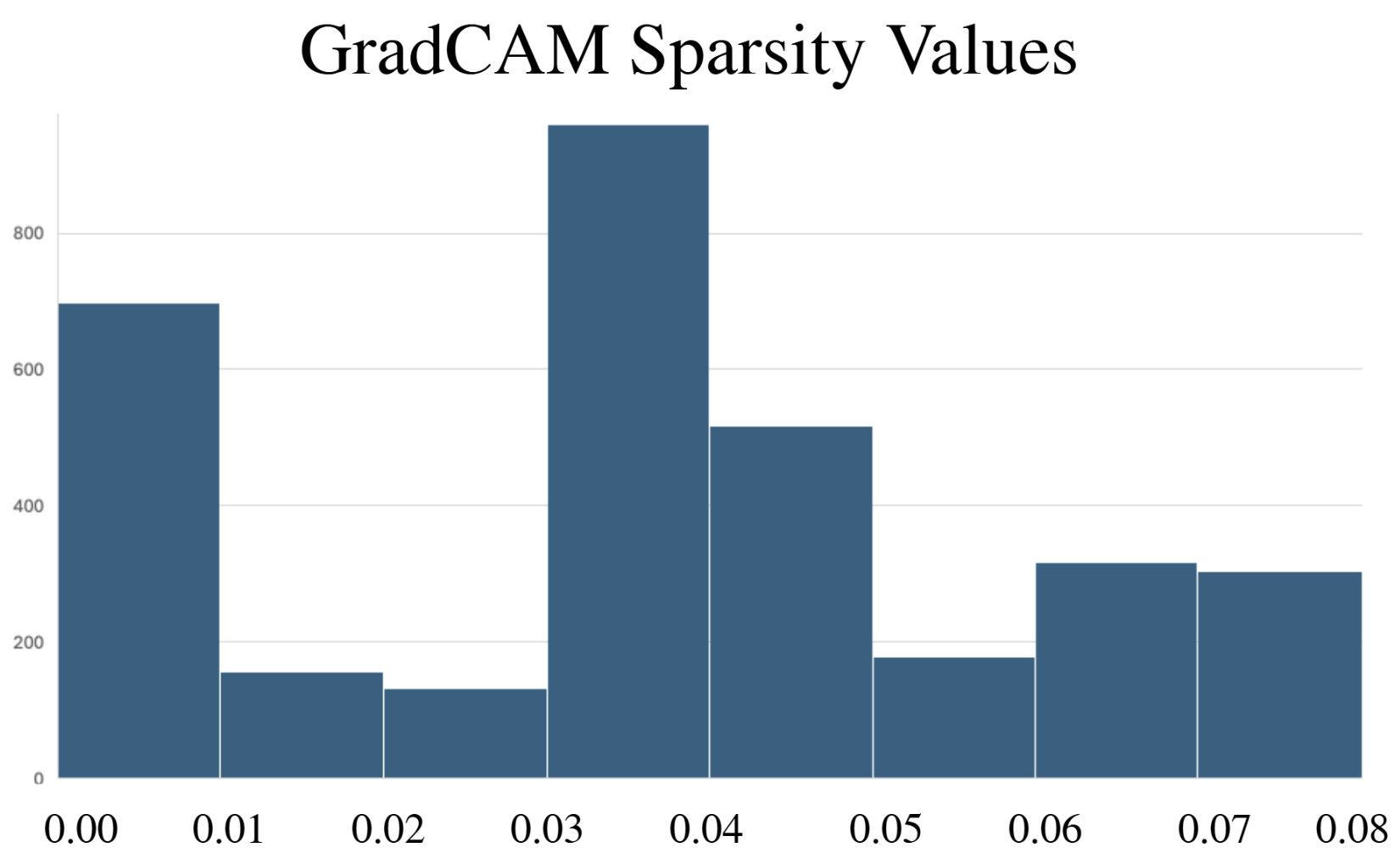}
  \end{minipage}
  \hfill
  \begin{minipage}[t]{0.32\linewidth}
    \centering
    \includegraphics[width=\linewidth]{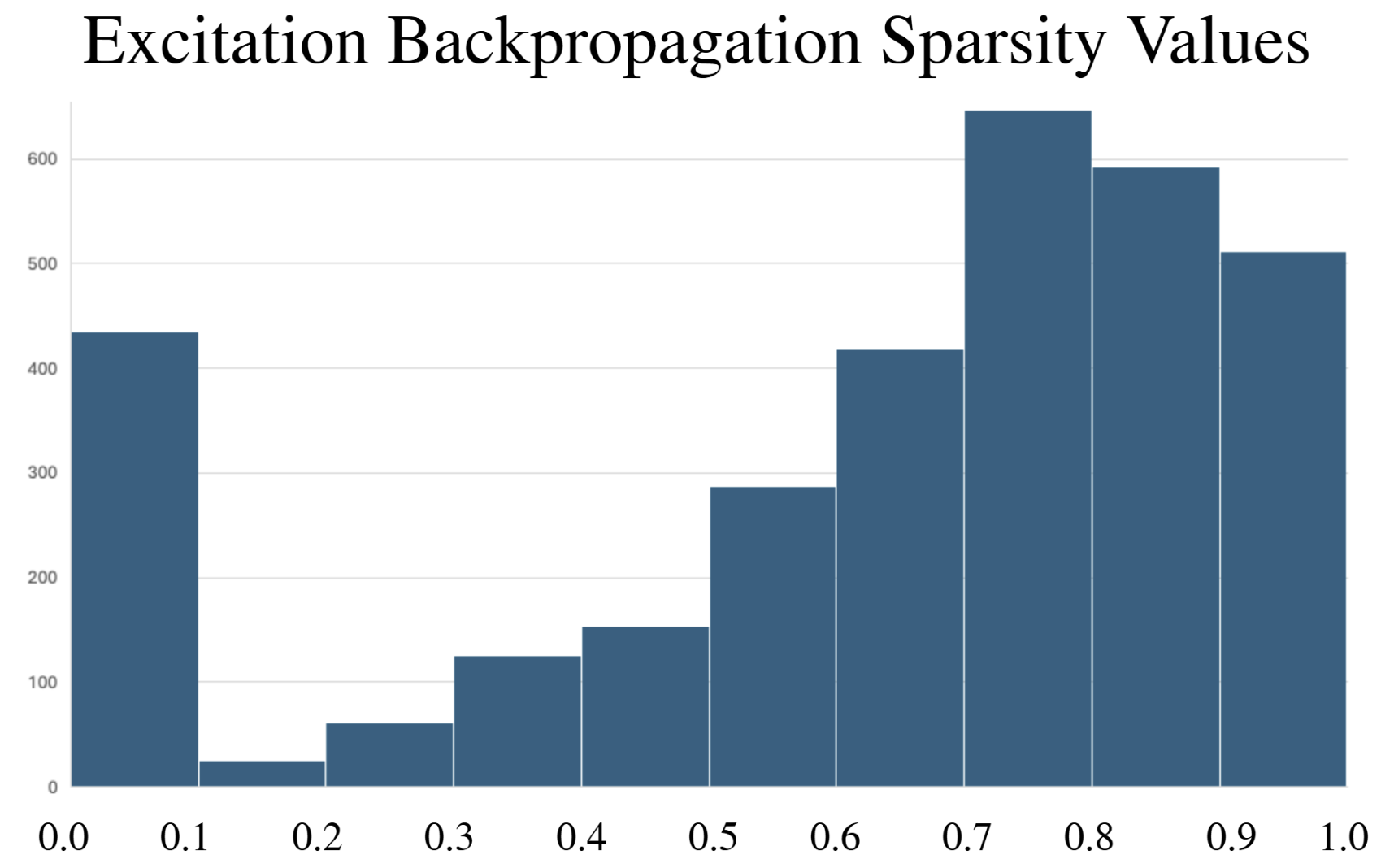}
  \end{minipage}
  \hfill
  \begin{minipage}[t]{0.32\linewidth}
    \centering
    \includegraphics[width=\linewidth]{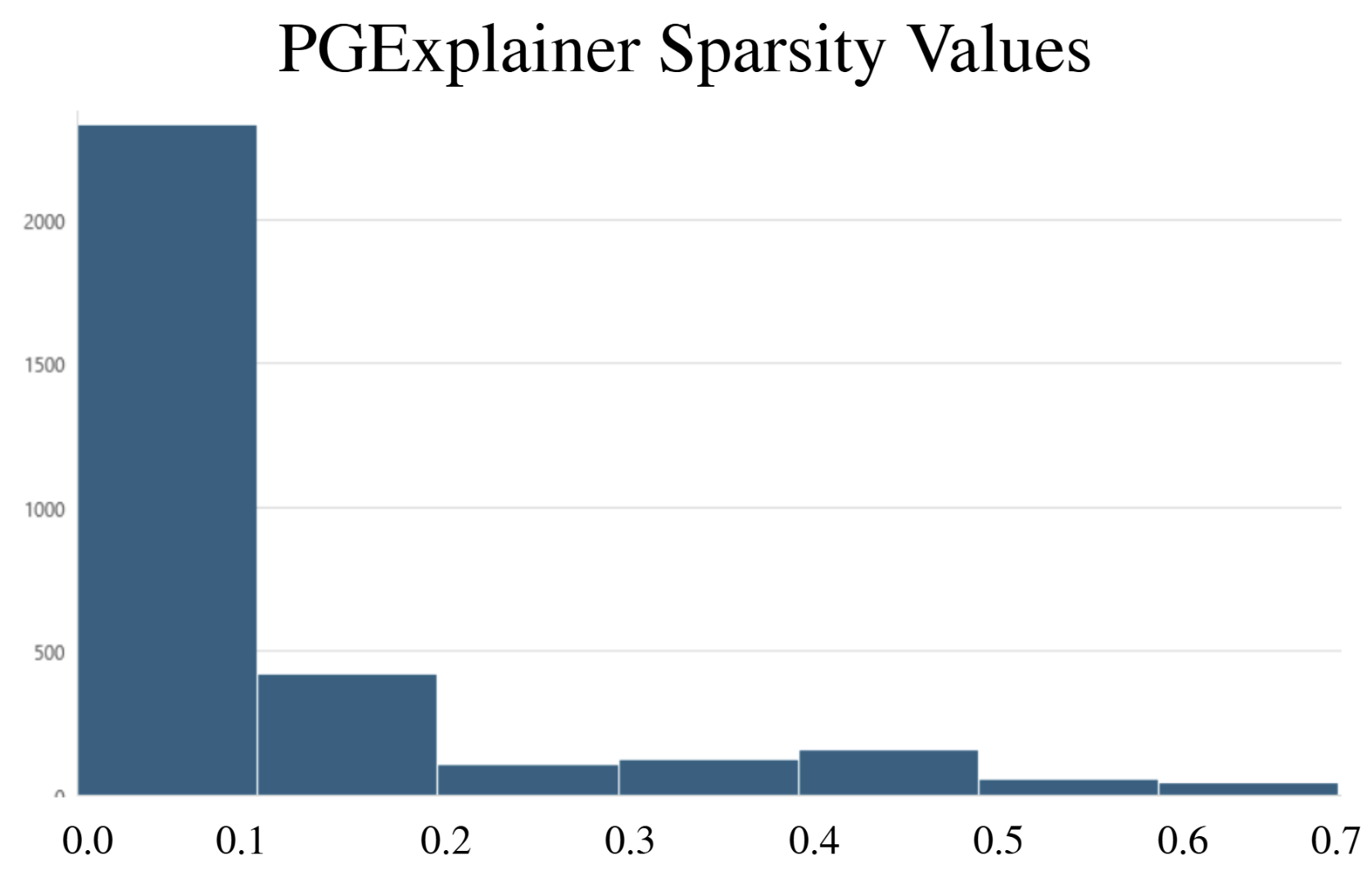}
  \end{minipage}

  \caption{Sparsity patterns across 124 binding proteins. A score of 0 reflects no sparsity, or that all amino acids were weighted in function prediction. A score of 1 reflects total sparsity, or that no amino acids were weighted in function prediction.}
  \label{fig:sparsity-comparison}
\end{figure}

By treating these three methods as an ensemble, residues flagged by all techniques become high-confidence hits, while those flagged by only one method highlight areas of epistemic uncertainty in the explanation itself.
  
\section{Conclusions and Next Steps}
Overall, our two case studies—the ARF6 GTPase and the lac repressor—paint a nuanced picture of DeepFRI’s strengths and limitations. On ARF6, all three explainability methods converged on biochemically critical residues and suggested high model fidelity in straightforward, conserved domains. In contrast, the lac repressor study revealed gaps: even after extensive mutagenesis-style perturbations, some functionally essential residues went undetected, highlighting that DeepFRI’s message-passing can overlook long-range or allosteric effects in more complex regulatory proteins.

A two-pronged strategy to improve interpretability and generalizability could be to guide experimental validation through targeted mutagenesis: begin with consensus residues in deep mutational scans, then probe method-specific sites via single-point substitutions  Second, DeepFRI’s input representations could benefit from fusing the GCN topology with per-residue embeddings from modern protein language models. This hybrid approach promises greater sensitivity to evolutionary and long-range signals without sacrificing structural context, ultimately yielding more trustworthy, actionable insights into protein function.

Our study emphasizes ways in which interpretability may not completely explain real world phenomena. By critically examining where explainability falls short, we pave the way for more reliable models that not only elucidate model decisions but also drive experimentally verifiable biological findings.
 


%
%
\bibliographystyle{splncs04}
\bibliography{main}

\end{document}